\documentstyle[12pt]{article}

\begin{document}
\begin{titlepage}
\begin{flushright}
{\sc SSCL-Preprint-492}\\
{\sc hep-ph/9308226}
\end{flushright}
\begin{center}
{\LARGE On a precise calculation of $(f_{B_s} / f_B) / (f_{D_s} / f_D)$\\
and its implications on the interpretation\\
of $B$ -- $\bar B$ mixing.}\\[.3in]
{\large Benjam\'\i n Grinstein\footnote{e-mail
address: grinstein@sscvx1.ssc.gov}\\
{Superconducting Super Collider Laboratory}\\
{Dallas, Texas  75237} }\\[.5in]

{\normalsize\bf Abstract}\\ [.2in]
\end{center}
{\small We observe that quantities like $(f_{B_{s}} / f_B) / (f_{D_{s}} /
f_D)$ are predicted to be unity {\em both} by heavy quark {\em and} by
light quark flavor symmetries.  Hence, the deviation from the symmetry
prediction must be simultaneously small in both symmetry breaking
parameters, {\em i.e.}, order of the ratio of light to heavy quarks
masses.  We estimate the size of the correction.  We observe that the
ratio of $(\Delta M/\Gamma)$ for $B_s - \bar B_s$ to $B - \bar B$
mixing can be expressed in terms of the measurable ratio $f_{D_{s}} /
f_D$ with good precision.  We comment on applications of these ideas
to other processes.}\\[.5in]
{\small August 4, 1993}
\end{titlepage}

Heavy quark symmetries\cite{isgur} have become an important tool in
the computation of decay widths and cross sections involving heavy
mesons and baryons (for reviews, see Ref.~\cite{MBWise}).  The
limitation of the method stems from the charm and bottom quark masses,
$m_c$ and $m_b$, being only slightly heavier than the typical hadronic
scale.  Depending on the particular quantity under investigation, the
corrections to heavy quark symmetry predictions can be more or less
sizable.  For example, there are many indications that corrections to
the scaling law for the pseudoscalar meson decay constant, $f_D\sim
1/\sqrt{M{_D}}$, are large, of order of 50\%\cite{Boucaud}.  And
deviations from the predicted value of the form factors for
semileptonic $B \to D$ and $B \to D^\ast$, and $ \Lambda _b \to
\Lambda _c$ transitions at zero recoil, are expected to be
small\cite{luke}.  It is of utmost importance to determine which
quantities are expected to be afflicted by large deviations from the
symmetry limit, and which are not.

In this letter we introduce a class of relations between observables
that arise as a result of heavy quark symmetry and separately as a
result of chiral symmetry of the light quarks.  Therefore, deviations
from these relations must be simultaneously small in the light quark
masses and in the inverse of the heavy quark masses, $1/m_Q$.  That
is, corrections are order $m_s / m_Q$.  This is to be contrasted with
the size of corrections to predictions from chiral symmetry alone,
$m_s / \Lambda$, or from heavy quark symmetries alone, $\Lambda /
m_Q$, with $\Lambda$ a typical hadronic scale.

To be specific consider the double ratio
\begin{equation}
R_1 = {{f_{B_{s}} / f_B}\over {f_{D_{s}} / f_D}}\ .
\end{equation}
Light quark flavor symmetries predict the ratio in the numerator and
the one in the denominator to deviate from unity by a quantity of
order $m_s / \Lambda$. Heavy quark flavor symmetries relate the
correction itself.  It can be expanded in powers of $\Lambda/m_Q$,
thus: ${m_s\over\Lambda} (a_0 + a_1 {\Lambda \over m_Q} +
\ldots)$, for some constants $a_i$.  The ratio (1) is therefore
\begin{equation}
R_1 = 1 + a_1 {\left({m_s\over m_b} - {m_s\over m_c}\right)}+ \cdots
\end{equation}
If we set, conservatively, $a_1 \approx 1$, and $m_s = 150\ MeV$, $m_c
= 1.5\ GeV$ and $m_b = 4.5\ GeV$, then the correction is only $R_1 - 1
\approx 7\%$.

Before we attempt a better estimate of the correction to $R_1 -1$,
let's see how this knowledge can be of use.  Of considerable interest
is the ratio
\begin{equation}
R_2 = {{(\Delta M / \Gamma)_{B_{s}}}\over {(\Delta M / \Gamma)_B}}~,
\end{equation}
which is a measurement of the relative strengths of $B_s - \bar B_s$
and $B - \bar B$ mixing.  In the standard model many uncertain
factors, such as top quark mass dependence, drop out in the ratio:
\begin{equation}
R_2 = \left| {V_{ts}\over V_{td}} \right|^2 \left({f_{B_{s}}\over f_B}
\right)^2 {B_{B_{s}}\over {B_{B}}}\ .
\end{equation}
To extract the fundamental ratio $\vert V_{ts} / V_{td} |$ from an
experimental determination of $R_2$, one needs knowledge of $f_{B_{s}}
/ f_B$ and the ratio of the mixing parameters $B_{B_{s}} / B_B$. Both
quantities have been studied recently\cite{grin,cho} using a
phenomenological lagrangian incorporating heavy quark and chiral
symmetries\cite{wise}.  The ratio $B_{B_{s}} / B_B$ is 1 with small
corrections, but the corrections to $(f_{B_{s}}/f_B)^2 -1$ are
sizable and cast a doubt on the reliability of this computation.
Clearly, a better approach is to use for $f_{B_{s }}/f_B$ the (soon to
be) measured ratio $f_{D_{s}}/f_D$. Therefore, to good approximation
\begin{equation}
R_2 = \left| {V_{ts}\over V_{td}} \right|^2 \left({f_{D_{s}}\over f_D}
\right)^2 ~.
\end{equation}

As an attempt to better estimate the corrections to $R_1 -1$ we use
the phenomenological lagrangian for heavy mesons and light
pseudoscalars that incorporates heavy and chiral symmetries.  We use
the lagrangian and notation of Ref.~\cite{grin}.  There are terms of
order $m _q / m_Q$ in the lagrangian and in the operator that
represents the axial current $\bar q
\gamma^{\mu} \gamma_5  Q$ in terms of effective fields.  Generally,
these involve unknown coefficients arising from strong interactions
dynamics.  But the coefficients of terms in the lagrangian responsible
for chiral $SU(3)$ symmetry breaking in the mass difference of heavy
vector and pseudoscalar mesons can be inferred from experiment.  Let
$\Delta^{(b)} \equiv M_{B^{\ast}} - M_B$ and $\Delta^{(b)}_s
\equiv M_{B^{\ast}_{s}} - M_{B_{s}}$ and define $\Delta ^{(c)}$ similarly
with $B$ replaced by $D$.  A class of computable corrections to $R_1
-1$, involving non-analytic dependence on the light quark masses, {\em
e.g.}, $\Delta\log m_s$, can be obtained by including the effects of
the mass shifts $\Delta$ in the renormalization of the heavy meson
fields. This is identical to the computation of $f_{D_{s}}/f_D$ of
Ref.~\cite{grin}, except that one must now keep track of the
dependence on $\Delta$, and neglect terms that cancel in the ratio
$R_1$.  It is also similar in spirit to the computation of $(1/m_c)^n$
effects of Ref.~\cite{randall}.  The result of the computation is
expressed in terms of the function
\begin{equation}
J (m, x) = (m^2 - 2 x ^2) \log (m^2 / \mu^2) - 4 x ^2 F (m / x)
\end{equation}
where
\begin{equation} F(y) \equiv \left\{ \begin{array}{rlllc}
{\sqrt{1 - y^2}} & \tanh^{-1} & \sqrt{1-y^2} & y \leq 1 \\
{- \sqrt{y^2 - 1}} & \tan^{-1} & \sqrt{y^2 - 1} & y \geq 1 & ~.
\end{array}
\right.
\end{equation}
The factor of $m^2 \log m^2$, which cancels in $R_1$, is retained for
comparison with the results of Ref.~\cite{grin}.  It is also useful to
introduce the chiral symmetry breaking parameters $\delta^{(b)} =
m_{B_{s}} -m_B$ and $\delta^{(c)}=m_{D_{s}}-m_D$.

Neglecting isospin breaking one obtains
\begin{eqnarray}
R_1 - 1  &\approx& {3g^{2}\over 32\pi^2f^2} \Biggl[ {2\over 3} J (M_{\eta},
\Delta^{(b)}_s ) + 2 J (M_K,
\Delta^{(b)} - \delta^{(b)})  \\
& & -{3\over 2} J (M_\pi, \Delta^{(b)}) - {1\over 6} J (M_{\eta},
\Delta^{(b)}) - J (M_K, \Delta^{(b)}_s +
\delta^{(b)} ) \Biggr]  - (b \to c) \nonumber
\end{eqnarray}

To estimate this numerically we take $g^2= 0.4$, its present
experimental upper bound\cite{barlay}, $f = f_K = 170\ MeV$,
$\Delta^{(b)}= \Delta{_s^{(b)}} = {1\over3} \Delta^{(c)} = {1\over3}
\Delta {_s^{(c)}} = 50\ MeV$, $\delta^{(b)}=\delta^{(c)}=150\ MeV$ and
the physical pseudoscalar masses. With these, the correction is $-
3.3\%$.  The deviation of each ratio, $f_{D_{s}}/f_D$ and
$f_{B_{s}}/f_B$, from unity is itself a factor of 4 larger.

We conclude by pointing out that there are many other symmetry
predictions of interest which are corrected at order $m_q / m_Q$ only.
An important example is the ratio of ratios of form factors for
semileptonic (or rare) $B$ and $D$ decays into light pseudoscalars
$(\pi$ and $K)$ or vector $(\rho$ and $K^\ast )$ mesons.  To be more
specific, let us parameterize the hadronic matrix elements in terms of
the Lorentz invariant form factors
\begin{eqnarray} <K(p_k) | \bar b\gamma^\mu s | B(p_B)> &=& f^{(B \to K)}_+
(p_B
+ p_K)^{\mu} +
f^{(B \to K)}_- (p_B - p_K)^{\mu}\\
<K (p_K) | \bar b \sigma^{\mu\nu}s | B (p_B) > &=& -2 i h^{(B\to K)} [p^\mu _B
p^\nu_K - p^\nu _B p^\mu_K] \nonumber
\end{eqnarray}
with analogous definitions for $f^{(B\to{\pi})}_{\pm}$ and
$h^{(B\to{\pi})}$ and for $D\to K$ and $D \to \pi$ matrix elements.
Writing the form factors as functions of $v
\cdot p_K = p_B \cdot p_K /m_B = p_D \cdot p_K/m_D$, one has the heavy quark
flavor symmetry relations
\begin{eqnarray}
& f^{(B \to K)}_+ / f^{(D \to K)}_+ &= \sqrt{m_b / m_c}\ ,\\
& f^{(B \to \pi )}_+ / f^{(D \to \pi)}_+ & = \sqrt{m_b / m_c}\ .\nonumber
\end{eqnarray}
This gives
\begin{equation}
{{f^{(B \to K)}_+ / f^{(B \to \pi)}_+} \over {f^{(D \to K)}_+ / f^{(D
\to \pi)}_+}} = 1\ .
\end{equation}
The same relation is obtained by an application of chiral symmetry to
the numerator and denominator separately.  Hence, the double ratio
deviates from unity in order $m_s / m_c$.  Analogous relations can be
written for $f_-$ and $h$.

Both form factors $f^{(D \to K)}_+$ and $f^{(D \to \pi)}_+$ are
experimentally accessible.  Computation of the rate for the rare decay
$B \to K \mu^+ \mu^-$ requires knowledge of the form factors $f^{(B
\to K)}_+$ and $h^{(B \to K)}$, while the rate for $B \to
\pi e \nu$ is expressed in terms of $f^{(B \to \pi)}_+$.
A precise determination of the rate for $B \to K \mu^+ \mu^-$
therefore requires measurement of $B \to \pi e \nu$ and an application
of heavy quark spin symmetries, which imply
\begin{equation}
2m_Q h = - f_- = f_+
\end{equation}
Note that one needs to invoke the use of heavy quark spin symmetry
only for B-meson form factor relations, and not for the lighter
D-meson.

The chiral symmetry prediction of the ratio of form
factors\cite{burdman} for $B \to K$ and $B \to
\pi$ is afflicted by rather large corrections\cite{falk}.  As we have seen, the
predictions of the ratio of ratios is expected to hold to much better
accuracy.  Therefore, a good measurement of $D \to Ke\nu$ and $D \to
\pi e\nu$ form factors could greatly aid in the interpretation of a
measurement of $B \to K \mu^+\mu^-$ and $B \to \pi e \nu$.

\vspace*{1in}
Acknowledgments.

Thanks to A. Kronfeld for conversations.  This work is support in part
by the Department of Energy under contract number DE-AC35-89ER40486.

\end{document}